\documentclass[a4paper,12pt]{article}
\usepackage[affil-it]{authblk}
\usepackage{enumitem}
\usepackage{amssymb}
\usepackage{amsmath}
\usepackage{graphicx}
\usepackage{multirow}

\newcommand{\ket}[1]{\ensuremath{\left|#1\right\rangle}}

\providecommand{\keywords}[1]
{
	\small    
	\textbf{\textit{Keywords---}} #1
}

\begin{document}
	
	\title{Improving the Security of ``Measurement-Device-Independent Quantum Communication without Encryption"}
	
	\author{Nayana Das%
		\thanks{Email address: \texttt{dasnayana92@gmail.com} }}
	\affil{Applied Statistics Unit, Indian Statistical Institute, Kolkata, India.}
	
	\author{Goutam Paul %
		\thanks{Email address: \texttt{goutam.paul@isical.ac.in}}}
	\affil{Cryptology and Security Research Unit, R. C. Bose Centre for Cryptology and Security, Indian Statistical Institute, Kolkata, India.\\}
	
	\date{}
	
	\maketitle
	
	\begin{abstract}
		Recently in 2018, Niu et al. proposed a measurement-device-independent quantum secure direct communication protocol using Einstein-Podolsky-Rosen pairs and generalized it to a quantum dialogue protocol (Niu et al., Science bulletin 63.20, 2018). By analyzing these protocols we find some security issues in both these protocols. In this work, we show that both the protocols are not secure against information leakage, and a third party can get half of the secret information without any active attack. We also propose suitable modifications of these protocols to improve the security.
	\end{abstract}
	
	\keywords{Cryptography; Information leakage; Information-theoretic security; Measurement Device Independent; Quantum secure direct communication; Quantum dialogue.}
	
	\section{Introduction}
	\label{intro}
	Quantum cryptography, where the security is based on the laws of quantum physics, is a remarkable application of quantum mechanics in the field of information theory. In 1984, Bennett and Brassard first used quantum resources to complete a cryptographic task, and they generated a secret key between two parties, which is the first quantum key distribution (QKD) protocol. This is called the BB84 protocol~\cite{bennett2020quantum}, where the parties use a sequence of single photons randomly prepared in the rectilinear basis ($\{\ket{0},\ket{1}\}$), and the diagonal basis ($\{\ket{+},\ket{-}\}$) to produce a random secret key. After that, various QKD protocols have been proposed by many researchers, such as Ekert's protocol~\cite{ekert1991quantum}, B92 protocol~\cite{bennett1992quantum1}, BBM92 protocol~\cite{bennett1992quantum}, SARG04 protocol~\cite{scarani2004quantum} and so on~\cite{long2002theoretically,xue2002conditional,deng2004bidirectional,lo2012measurement}. 
	
	In 2002, quantum secure direct communication (QSDC), a new concept of communicating messages securely over a quantum channel without any shared key, was first proposed by Long et. al.~\cite{long2002theoretically}. This is a process of secure communication without any cryptographic encryption or decryption. Here the sender encodes the message on some qubits by using some predefined encoding rule and sends these qubits to the receiver through a quantum channel. From its initial stage, QSDC has drawn a lot of attention and has become an interesting topic of research~\cite{beige2001secure,bostrom2002ping,deng2003two,deng2004secure,wang2005quantum,wang2005multi,wang2006quantum,zhang2017quantum,das2021quantum}. A bidirectional QSDC protocol, called quantum dialogue (QD), was first proposed by Nguyen in 2004~\cite{nguyen2004quantum}. Now there is a large collection of QD protocols, for example Refs.~\cite{zhang2004deterministic,zhong2005quantum,xia2006quantum,xin2006secure,gao2010two,maitra2017measurement,das2020two}. QSDC protocols for three or more parties are discussed in~\cite{gao2005deterministic,jin2006three,ting2005simultaneous,tan2014multi,das2021secure}.

	Recently, Niu et al. proposed a measurement-device-independent (MDI) QSDC protocol using Einstein-Podolsky-Rosen (EPR) pairs~\cite{niu2018measurement}. Then they generalized this one-way communication to a bidirectional one and proposed an MDI-QD protocol. In their protocols, the two legitimate parties prepare two sets of EPR pairs in their place, and send the partner qubits of their EPR pairs to an untrusted third party, since the condition for being an MDI protocol is that, all the measurements during the communication process should be performed by an untrusted third party (who may be an eavesdropper). Here we analyze these protocols and point out that the secret messages are not transmitted securely for both the protocols. We show that fifty percent of the information about the secret message bits is leaked out in both the protocols. In other words, in the perspective of information theory and cryptography, these protocols are not secure. This type of security loophole of information leakage in various QSDC and QD protocols are discussed in~\cite{zhong2007improvement,gao2008comment,gao2008revisiting,tan2008classical,fei2008teleportation,wang2011information,gao2014information,das2020cryptanalysis}. We also propose modifications of these protocols to improve their security.
	
	The rest of the paper is organized as follows. In the next section, we briefly describe the MDI-QSDC and MDI-QD protocols proposed by Niu et al.~\cite{niu2018measurement}. In Section~\ref{sec3:analysis}, we analyze the security loophole of the above protocols, and then our proposed remedy is given in Section~\ref{sec4:modified protocol}. Finally, Section~\ref{conclusion} concludes our work.
	
	\section{Brief Review of Niu et al.'s Protocols \cite{niu2018measurement}}
	\label{sec:review}
	In this section, we briefly describe the MDI-QSDC and MDI-QD protocols proposed by Niu et al. in 2018. 
	\subsection{MDI-QSDC protocol }\label{MDI-QSDC protocol}
	There are three parties in this protocol, namely, Alice, Bob and Charlie, where Alice wants to send some message to Bob, and Charlie is an untrusted third party, who performs all the measurements. They use the EPR pairs $\ket{\Phi^{+}}, \ket{\Phi^{-}}, \ket{\Psi^{+}}, \ket{\Psi^{-}}$ for sending the message bits, where,
	\begin{equation}
	\ket{\Phi^{\pm}}=\frac{1}{\sqrt{2}}(\ket{00} \pm \ket{11}),\\
	\ket{\Psi^{\pm}}=\frac{1}{\sqrt{2}}(\ket{01} \pm \ket{10}).
	\end{equation}
	The steps of the protocol are as follows: 
	\begin{enumerate}
		\item  \label{step1}Alice prepares $n$ EPR pairs randomly in $\ket{\Psi^{+}}$ and $\ket{\Psi^{-}}$ states and creates two sequences $S_{A_1}$ and $S_{A_2}$ of single photons, such that for $1 \leq i \leq n$, the $i$-th qubits of $S_{A_1}$ and $S_{A_2}$ are partners of each other in the $i$-th EPR pair. Similarly, Bob also prepares $S_{B_1}$ and $S_{B_2}$ from his $n$ EPR pairs randomly chosen from $\ket{\Psi^{+}}$ and $\ket{\Psi^{-}}$. Alice (Bob) also chooses $m$ single qubit states randomly from $\{\ket{0},\ket{1},\ket{+}=\frac{1}{\sqrt{2}}(\ket{0}+\ket{1}),\ket{-}=\frac{1}{\sqrt{2}}(\ket{0}-\ket{1})\}$ and inserts these qubits in random positions of $S_{A_2}$ ($S_{B_2}$), and let the new sequence be $C_{A_2}$ ($C_{B_2}$) containing $(n+m)$ single qubit states. 
		
		\item  \label{step2}Alice (Bob) sends the sequence $C_{A_2}$ ($C_{B_2}$) to Charlie and keeps $S_{A_1}$ ($S_{B_1}$) in her (his) lab.
		
		\item Charlie makes Bell measurement on each pair of $C_{A_2}$ and $C_{B_2}$ (i.e., the $i$-th Bell measurement on the $i$-th qubit of $C_{A_2}$ and the $i$-th qubit of $C_{B_2}$, $1\leq i \leq n+m$) and announces the results. \label{1st_measure}
		
		\item  \label{step4}Alice and Bob announce the positions of the single qubit states in the sequences $C_{A_2}$ and $C_{B_2}$ respectively. For $1\leq i \leq n+m$, four cases may arise.
		\begin{enumerate}
			\item  \label{after_1st_measure}If the $i$-th qubit of $C_{A_2}$ and the $i$-th qubit of $C_{B_2}$ are from $S_{A_2}$ and $S_{B_2}$ respectively, then as a result of quantum entanglement swapping~\cite{zukowski1993event}, the Bell measurement causes the corresponding partner qubits of $S_{A_1}$ and $S_{B_1}$ become an EPR pair, which is shown in Equation~\eqref{ent_swp}.
			\begin{equation}
			\label{ent_swp}
			\begin{aligned}
			\ket{\Psi^{+}}_{A_1A_2}\ket{\Psi^{+}}_{B_1B_2}={} & \frac{1}{2}(\ket{\Psi^{+}}_{A_1B_1}\ket{\Psi^{+}}_{A_2B_2}-\ket{\Psi^{-}}_{A_1B_1}\ket{\Psi^{-}}_{A_2B_2}+\\
			& \ket{\Phi^{+}}_{A_1B_1}\ket{\Phi^{+}}_{A_2B_2}-\ket{\Phi^{-}}_{A_1B_1}\ket{\Phi^{-}}_{A_2B_2}),\\
			\ket{\Psi^{-}}_{A_1A_2}\ket{\Psi^{+}}_{B_1B_2}={} & \frac{1}{2}(\ket{\Psi^{-}}_{A_1B_1}\ket{\Psi^{+}}_{A_2B_2}-\ket{\Psi^{+}}_{A_1B_1}\ket{\Psi^{-}}_{A_2B_2}+\\
			& \ket{\Phi^{-}}_{A_1B_1}\ket{\Phi^{+}}_{A_2B_2}-\ket{\Phi^{+}}_{A_1B_1}\ket{\Phi^{-}}_{A_2B_2}),\\
			\ket{\Psi^{+}}_{A_1A_2}\ket{\Psi^{-}}_{B_1B_2}={} & \frac{1}{2}(\ket{\Psi^{+}}_{A_1B_1}\ket{\Psi^{-}}_{A_2B_2}-\ket{\Psi^{-}}_{A_1B_1}\ket{\Psi^{+}}_{A_2B_2}+\\
			& \ket{\Phi^{-}}_{A_1B_1}\ket{\Phi^{+}}_{A_2B_2}-\ket{\Phi^{+}}_{A_1B_1}\ket{\Phi^{-}}_{A_2B_2}),\\
			\ket{\Psi^{-}}_{A_1A_2}\ket{\Psi^{-}}_{B_1B_2}={} & \frac{1}{2}(\ket{\Psi^{-}}_{A_1B_1}\ket{\Psi^{-}}_{A_2B_2}-\ket{\Psi^{+}}_{A_1B_1}\ket{\Psi^{+}}_{A_2B_2}+\\
			& \ket{\Phi^{+}}_{A_1B_1}\ket{\Phi^{+}}_{A_2B_2}-\ket{\Phi^{-}}_{A_1B_1}\ket{\Phi^{-}}_{A_2B_2}).\\
			\end{aligned}
			\end{equation}
			
			\item If the $i$-th qubit of $C_{A_2}$ is from $S_{A_2}$ and the $i$-th qubit of $C_{B_2}$ is any single qubit from the set $\{\ket{0},\ket{1},\ket{+},\ket{-}\}$, then Alice and Bob discard the $i$-th Bell measurement result.
			
			\item If the $i$-th qubit of $C_{A_2}$ is a single qubit from the set $\{\ket{0},\ket{1},\ket{+},\ket{-}\}$ and the $i$-th qubit of $C_{B_2}$ is from $S_{B_2}$, then also Alice and Bob discard the $i$-th Bell measurement result.
			
			\item If both the $i$-th qubits of $C_{A_2}$ and $C_{B_2}$ are from  the set $\{\ket{0},\ket{1},\ket{+},\ket{-}\}$, then Alice and Bob exchange the basis information of their single qubits. If the bases are different, then they discard the $i$-th Bell measurement result. Else it is used for security checking. A pair of single qubits with identical bases can be written as: 
			\begin{equation}
			\label{sinle_bell}
			\begin{aligned}
			\ket{0}_{A_2}\ket{0}_{B_2}=    \frac{1}{\sqrt{2}}(\ket{\Phi^{+}}_{A_2B_2}+\ket{\Phi^{-}}_{A_2B_2}),    \\
			\ket{1}_{A_2}\ket{1}_{B_2}=    \frac{1}{\sqrt{2}}(\ket{\Phi^{+}}_{A_2B_2}-\ket{\Phi^{-}}_{A_2B_2}),    \\
			\ket{0}_{A_2}\ket{1}_{B_2}=    \frac{1}{\sqrt{2}}(\ket{\Psi^{+}}_{A_2B_2}+\ket{\Psi^{-}}_{A_2B_2}),    \\
			\ket{1}_{A_2}\ket{0}_{B_2}=    \frac{1}{\sqrt{2}}(\ket{\Psi^{+}}_{A_2B_2}-\ket{\Psi^{-}}_{A_2B_2});\\        
			\end{aligned}
			\end{equation}
			and
			\begin{equation}
			\label{sinle_bell_+}
			\begin{aligned}
			\ket{+}_{A_2}\ket{+}_{B_2}=    \frac{1}{\sqrt{2}}(\ket{\Phi^{+}}_{A_2B_2}+\ket{\Psi^{+}}_{A_2B_2}),    \\
			\ket{-}_{A_2}\ket{-}_{B_2}=    \frac{1}{\sqrt{2}}(\ket{\Phi^{+}}_{A_2B_2}-\ket{\Psi^{+}}_{A_2B_2}).\\        
			\ket{+}_{A_2}\ket{-}_{B_2}=    \frac{1}{\sqrt{2}}(\ket{\Phi^{-}}_{A_2B_2}-\ket{\Psi^{-}}_{A_2B_2}),    \\
			\ket{-}_{A_2}\ket{+}_{B_2}=    \frac{1}{\sqrt{2}}(\ket{\Phi^{-}}_{A_2B_2}+\ket{\Psi^{-}}_{A_2B_2}).    \\
			\end{aligned}
			\end{equation}
			
			Using the relations~\eqref{sinle_bell} and~\eqref{sinle_bell_+}, Alice and Bob estimate the error in the channel and decide to continue the protocol or not.
		\end{enumerate}
		
		\item  \label{step5}Alice and Bob discard the qubits, which are not entangled, from their sequences $S_{A_1}$ and $S_{B_1}$, and make the new sequences $M_A$ and $M_B$ respectively. Let the number of discarded qubits from each set be $\delta$,  and then each new sequence contains $(n-\delta)$ single qubits. Alice performs the unitary operation $\sigma_z$~\cite{nielsen2002quantum}, on the qubits of $M_A$, whose initial states were $\ket{\Psi^{+}}$. This process is equivalent to the fact that Alice prepared all the initial EPR pairs in $\ket{\Psi^{-}}$ state. Now, only Bob knows the actual state of the qubit pairs $({M_A}_i,{M_B}_i)$ for $1 \leq i \leq n-\delta$, where ${M_A}_i$ and ${M_B}_i$ are the $i$-th qubits of the sequences $M_A$ and $M_B$ respectively. Due to quantum entanglement swapping, $({M_A}_i,{M_B}_i)$ is in a Bell state (see Equation~\eqref{ent_swp}). 
		
		\item  \label{step6}Message encoding: Alice puts some random checking bits on random positions of her message. She applies one of the four unitary operators (Pauli matrices~\cite{nielsen2002quantum}), $I$, $\sigma_x$, $i\sigma_y$ and $\sigma_{z}$, on the qubits of $M_A$, to encode the information $00$, $01$, $10$, and $11$ respectively. To make the protocol secure against the intercept-and-resend attack, Bob randomly applies $I$ or $\sigma_{z}$ on the qubits of $M_B$.
		
		\item Alice (Bob) sends the sequence $M_{A}$ ($M_{B}$) to Charlie, who measures each pair of qubits of $M_A$ and $M_B$ on Bell basis and announces the results. From the measurement results, Bob decodes the message of Alice. Then Alice announces the positions and value of the random checking bits, and from this information, they can check the integrity of the message. A non-negligible error implies the existence of some eavesdropper in the channel. \label{2nd_measure}
	\end{enumerate}
	
	\subsection{MDI-QD protocol }
	This is a simple generalization of the previous MDI-QSDC protocol. The first five steps are the same as above. To encode their messages, Alice and Bob divide the pair of sequence $(M_A,M_B)$ into two disjoint parts $(M_{A}^1,M_{B}^1)$ and $(M_{A}^2,M_{B}^2)$. One part is used for sending the message from Alice to Bob and another part is used for sending a message from Bob to Alice.
	
	\section{Security loophole of the MDI-QSDC protocol~\cite{niu2018measurement}}\label{sec3:analysis}
	In this section, we explicitly analyze the above MDI-QSDC protocol discussed in Section~\ref{MDI-QSDC protocol}. After Charlie has done the first set of Bell measurements of the qubits pairs of $S_{A_2}$ and $S_{B_2}$ in Step~\ref{1st_measure}, the qubits pairs of $S_{A_1}$ and $S_{B_1}$ become entangled due to entanglement swapping (Step~\ref{after_1st_measure}). Now from Equation~\eqref{ent_swp}, we can see that, if the Bell measurement results of the qubits pairs of $S_{A_2}$ and $S_{B_2}$ are $\ket{\Phi^{+}}_{A_2B_2}$ or $\ket{\Phi^{-}}_{A_2B_2}$, then also the states of the qubit pairs of $S_{A_1}$ and $S_{B_1}$ are $\ket{\Phi^{+}}_{A_1B_1}$ or $\ket{\Phi^{-}}_{A_1B_1}$. Similarly, the state of the qubit pair $(A_2,B_2)=\ket{\Psi^{\pm}}_{A_2B_2}$ implies the state of the qubit pair $(A_1,B_1)=\ket{\Psi^{\pm}}_{A_1B_1}$ or $\ket{\Psi^{\mp}}_{A_1B_1}$.
	
	After security checking, Alice and Bob discard the qubits, which are not entangled, from their sequences $S_{A_1}$ and $S_{B_1}$, and make the new sequences $M_A$ and $M_B$ respectively.  So, from the Bell measurement results of the qubit pairs $(A_2,B_2)$, Charlie knows the states of the qubit pairs $(A_1,B_1)$, are either $\ket{\Phi^{\pm}}_{A_1B_1}$ or $\ket{\Psi^{\pm}}_{A_1B_1}$. That is, for $1 \leq i \leq n-\delta$, Charlie exactly knows that the qubit pairs $({M_A}_i,{M_B}_i)$ are in set $\varPhi=\{\ket{\Phi^{+}},\ket{\Phi^{-}}\}$ or in set $\varPsi=\{\ket{\Psi^{+}},\ket{\Psi^{-}}\}$.
	
	Now Alice applies $\sigma_z$ on the qubits of $M_A$, whose corresponding initial states were $\ket{\Psi^{+}}$. It is easy to check that, if Alice applies $\sigma_z$ on $M_{A_i}$ for some $i$, then the state of the qubit pair $({M_A}_i,{M_B}_i)$ changes from either $\ket{\Phi^{\pm}}$ to $\ket{\Phi^{\mp}}$ or $\ket{\Psi^{\pm}}$ to $\ket{\Psi^{\mp}}$. Thus Charlie's knowledge about the state of $({M_A}_i,{M_B}_i)$ remains same.
	
	Then  Alice encodes her message on the qubits of $M_A$ by using the unitary operations  $I$, $\sigma_x$, $i\sigma_y$ and $\sigma_{z}$ corresponding the message bits $00$, $01$, $10$, and $11$ respectively. That is, the unitary operators $I$ and $\sigma_z$ are used to encode the message bits $bb$, and the unitary operators $\sigma_x$ and $i\sigma_y$ are used to encode the message bits $b\bar{b}$, where $b \in \{0,1\}$ and $\bar{b}$ = bit complement of $b$. Bob also randomly applies $I$ or $\sigma_{z}$ on the qubits of $M_B$.
	They send $M_A$ and $M_B$ to Charlie, who measures each pair of qubits $({M_A}_i,{M_B}_i)$ in Bell basis, and announces the results. All the different cases are given in Table~\ref{cases of qsdc}.
	
	\begin{table}[]
		\centering
		\renewcommand*{\arraystretch}{1.7}
		\caption{Different cases of MDI-QSDC~\cite{niu2018measurement}.}
		\setlength{\tabcolsep}{10pt}
		\resizebox{1.05\textwidth}{!}{
			\begin{tabular}{|c|c|c|c|c|}
				\hline
				\textbf{State of $\mathbf{(M_{A_i},M_{B_i})}$}    & \textbf{Message bits} & \textbf{Alice's unitary}     & \textbf{Bob's unitary} & \textbf{State of $\mathbf{(M_{A_i},M_{B_i})}$} \\ 
				\textbf{ before encoding}    & \textbf{of Alice} & \textbf{operation on $\mathbf{M_{A_i}}$}     & \textbf{operation on $\mathbf{M_{B_i}}$} & \textbf{after encoding} \\ 
				\hline
				\multirow{8}{*}{$\ket{\Phi^{+}}$} & \multirow{2}{*}{00}   & \multirow{2}{*}{$I$}         & $I$                    & $\ket{\Phi^{+}}$                 \\ \cline{4-5} 
				&                       &                              & $\sigma_z$             & $\ket{\Phi^{-}}$                 \\ \cline{2-5} 
				& \multirow{2}{*}{01}   & \multirow{2}{*}{$\sigma_x$}  & $I$                    & $\ket{\Psi^{+}}$                 \\ \cline{4-5} 
				&                       &                              & $\sigma_z$             & $\ket{\Psi^{-}}$                 \\ \cline{2-5} 
				& \multirow{2}{*}{10}   & \multirow{2}{*}{$i\sigma_y$} & $I$                    & $\ket{\Psi^{-}}$                 \\ \cline{4-5} 
				&                       &                              & $\sigma_z$             & $\ket{\Psi^{+}}$                 \\ \cline{2-5} 
				& \multirow{2}{*}{11}   & \multirow{2}{*}{$\sigma_z$}  & $I$                    & $\ket{\Phi^{-}}$                 \\ \cline{4-5} 
				&                       &                              & $\sigma_z$             & $\ket{\Phi^{+}}$                 \\ \hline
				\multirow{8}{*}{$\ket{\Phi^{-}}$} & \multirow{2}{*}{00}   & \multirow{2}{*}{$I$}         & $I$                    & $\ket{\Phi^{-}}$                 \\ \cline{4-5} 
				&                       &                              & $\sigma_z$             & $\ket{\Phi^{+}}$                 \\ \cline{2-5} 
				& \multirow{2}{*}{01}   & \multirow{2}{*}{$\sigma_x$}  & $I$                    & $\ket{\Psi^{-}}$                 \\ \cline{4-5} 
				&                       &                              & $\sigma_z$             & $\ket{\Psi^{+}}$                 \\ \cline{2-5} 
				& \multirow{2}{*}{10}   & \multirow{2}{*}{$i\sigma_y$} & $I$                    & $\ket{\Psi^{+}}$                 \\ \cline{4-5} 
				&                       &                              & $\sigma_z$             & $\ket{\Psi^{-}}$                 \\ \cline{2-5} 
				& \multirow{2}{*}{11}   & \multirow{2}{*}{$\sigma_z$}  & $I$                    & $\ket{\Phi^{+}}$                 \\ \cline{4-5} 
				&                       &                              & $\sigma_z$             & $\ket{\Phi^{-}}$                 \\ \hline
				\multirow{8}{*}{$\ket{\Psi^{+}}$} & \multirow{2}{*}{00}   & \multirow{2}{*}{$I$}         & $I$                    & $\ket{\Psi^{+}}$                 \\ \cline{4-5} 
				&                       &                              & $\sigma_z$             & $\ket{\Psi^{-}}$                 \\ \cline{2-5} 
				& \multirow{2}{*}{01}   & \multirow{2}{*}{$\sigma_x$}  & $I$                    & $\ket{\Phi^{+}}$                 \\ \cline{4-5} 
				&                       &                              & $\sigma_z$             & $\ket{\Phi^{-}}$                 \\ \cline{2-5} 
				& \multirow{2}{*}{10}   & \multirow{2}{*}{$i\sigma_y$} & $I$                    & $\ket{\Phi^{-}}$                 \\ \cline{4-5} 
				&                       &                              & $\sigma_z$             & $\ket{\Phi^{+}}$                 \\ \cline{2-5} 
				& \multirow{2}{*}{11}   & \multirow{2}{*}{$\sigma_z$}  & $I$                    & $\ket{\Psi^{-}}$                 \\ \cline{4-5} 
				&                       &                              & $\sigma_z$             & $\ket{\Psi^{+}}$                 \\ \hline
				\multirow{8}{*}{$\ket{\Psi^{-}}$} & \multirow{2}{*}{00}   & \multirow{2}{*}{$I$}         & $I$                    & $\ket{\Psi^{-}}$                 \\ \cline{4-5} 
				&                       &                              & $\sigma_z$             & $\ket{\Psi^{+}}$                 \\ \cline{2-5} 
				& \multirow{2}{*}{01}   & \multirow{2}{*}{$\sigma_x$}  & $I$                    & $\ket{\Phi^{-}}$                 \\ \cline{4-5} 
				&                       &                              & $\sigma_z$             & $\ket{\Phi^{+}}$                 \\ \cline{2-5} 
				& \multirow{2}{*}{10}   & \multirow{2}{*}{$i\sigma_y$} & $I$                    & $\ket{\Phi^{+}}$                 \\ \cline{4-5}

				&                       &                              & $\sigma_z$             & $\ket{\Phi^{-}}$                 \\ \cline{2-5} 
				& \multirow{2}{*}{11}   & \multirow{2}{*}{$\sigma_z$}  & $I$                    & $\ket{\Psi^{+}}$                 \\ \cline{4-5} 
				&                       &                              & $\sigma_z$             & $\ket{\Psi^{-}}$                 \\ \hline
			\end{tabular}
		}
		\label{cases of qsdc}
	\end{table}
	
	We now show that, in the MDI-QSDC protocol~\cite{niu2018measurement}, the untrusted third party Charlie (or any eavesdropper) can get partial information about the secret without any active attack. For this, we need to discuss the effects of the encoding rules in this MDI-QSDC protocol. Without loss of generality, suppose the joint state of ${M_A}_i$, ${M_B}_i$ before encoding is $\ket{\Phi^{+}}$, then Charlie knows that the joint state is in the set $\varPhi$.
	
	After Charlie measures $({M_A}_i,{M_B}_i)$ in Bell basis, if the measurement result is in  the set $\varPhi$, then from Table~\ref{cases of qsdc}, Charlie concludes that, the secret information is either $00$ or $11$. Again if the measurement result is in the set $\varPsi$, then from Table~\ref{cases of qsdc}, Charlie concludes that, the secret information is either $01$ or $10$. Similarly, for the other cases, Charlie exactly knows that the secret information is $bb$ or $b\bar{b}$. For both the cases, Charlie can get the exact secret information with probability $1/2$, thus the Shannon entropy, which measures the amount of uncertainty, is equal to $-\sum_{j=1}^{2}\frac{1}{2}\log\frac{1}{2}=1$ bit. That means, only one bit among two bits of secret information is unknown to Charlie. One may note that, from the viewpoint of information theory, this is equivalent to the event that, among two bits of secret information, Charlie knows the exact value of one bit and does not have any knowledge about the other bit. Thus we can say that, here in this MDI-QSDC protocol, only fifty percent of the secret message communicated securely. 
	
	By the same argument, we can say that the MDI-QD protocol proposed in~\cite{niu2018measurement} is also not secure against information leakage, and in this protocol, only fifty percent of the secret messages communicated securely.
	
	Now, we find the root of this information leakage problem in these protocols. Let for some~$i$, $M_{A_i} \in M_A$ and $M_{B_i} \in M_B$, and after Alice and Bob apply their unitary operators, the states $M_{A_i}$ and $M_{B_i}$ become $N_{A_i}$ and $N_{B_i}$ respectively. If the joint state $(M_{A_i},M_{B_i})\in \varPhi$ or $\varPsi$, then after applying $I$ or $\sigma_z$ on $M_{A_i}$ (or $M_{B_i}$),  the joint state $(N_{A_i},M_{B_i})$ (or $(M_{A_i},N_{B_i})$) remains in the same set $\varPhi$ or $\varPsi$ respectively. In other words, both $I$ and $\sigma_z$ are applied on $M_{A_i}$ or $M_{B_i}$ or both $M_{A_i}$ and $M_{B_i}$, map the set $\varPhi$  to $\varPhi$, and $\varPsi$ to $\varPsi$.
	That is, for both the mappings, the domain and the range sets are same, and if both the joint states $(M_{A_i},M_{B_i})$ and $(N_{A_i},N_{B_i})$ belong to the same subset of the Bell states $\varPhi$ or $\varPsi$, then Charlie concludes that the message bits are $bb$. Otherwise, when $(M_{A_i},M_{B_i})$ and $(N_{A_i},N_{B_i})$ belong to two different subsets $\varPhi$ or $\varPsi$, then Charlie concludes that the message bits are $b\bar{b}$ (i.e., Alice applies  $\sigma_x$ or $i\sigma_y$ on $M_{A_i}$). So, the main problem in this encoding rule is, Bob's random unitary operations can not lower down the information of Charlie about the secret message. In the next section, we propose a remedy to overcome this security flaw.

	\section{Proposed modification of MDI-QSDC protocol}\label{sec4:modified protocol}
	In this section, we modify the MDI-QSDC protocol, to make it secure against information leakage. To resolve the problem discussed in Section~\ref{sec3:analysis}, Bob needs to apply some random unitary operators on $M_{B_i}$ such that the the union of the range sets, of his unitary operators, becomes the whole set of Bell states, i.e., for each $(M_{A_i},M_{B_i}) \in \varPhi$ or $\varPsi$ and $(N_{A_i},N_{B_i}) \in \varPhi \cup \varPsi$, there exist all the four possibilities of Alice's two bits message $b_1b_2$  ($b_1,b_2\in \{0,1\}$).
	
	The modified protocol is almost same as the original one. In our modified MDI-QSDC protocol, Step~\ref{step1} to Step~\ref{step5} and Step~\ref{2nd_measure} are same as the MDI-QSDC protocol discussed in Section~\ref{MDI-QSDC protocol}. In Step 6, the encoding process of Alice is the same as the previous one, and Bob randomly applies $\sigma_x$ and $I$ on the qubits of $M_B$ (instead of $\sigma_z$ and $I$ in the original one). All the different cases, of the states of the qubit pairs of $M_A$ and $M_B$, before and after encoding are given in Table~\ref{mod_table}. 
	
	We will now show that this modified protocol is secure against information leakage. Again without loss of generality, suppose the joint state of ${M_A}_i$, ${M_B}_i$ before encoding is $\ket{\Phi^{+}}$, then Charlie knows that the joint state is either $\ket{\Phi^{+}}$ or $\ket{\Phi^{-}}$. From Table~\ref{mod_table}, it is easy to check that, before encoding, if the joint state is $\ket{\Phi^{\pm}}$, then all the four Bell states can arise after encoding any two message bits $b_1b_2$. Thus Charlie's knowledge, about the joint state before encoding, does not help him to extract any information about the secret bits. Similarly for the other cases also Charlie can not get any secret information about the message bits.
	
	We can also modify the MDI-QD protocol of \cite{niu2018measurement}, with a similar approach, i.e., the receiver applies the unitary $I$ and $\sigma_x$ randomly on his (her) state at the time of encoding.
	
	\subsection{Other Pauli operators to fix the issue}
	One can ask, what happen if Bob chooses any other pair of Pauli matrices as his random unitary operators. To check this, we consider two sets of linear transformations $\mathcal{F}_1=\{I, \sigma_z\}$ and $\mathcal{F}_2=\{\sigma_x, i\sigma_y\}$ (note that, every matrix is a linear transformation), where both the domain and range of these linear transformations are $\varPhi$ and $\varPsi$. Then, $f \in \mathcal{F}_1$ implies that $f$ maps the set $\varPhi$ to $\varPhi$ and the set $\varPsi$ to $\varPsi$ (ignoring the global phase of the Bell states). Again, $f \in \mathcal{F}_2$ implies that $f$ maps the set $\varPhi$ to $\varPsi$ and the set $\varPsi$ to $\varPhi$. Let for any mapping $f$, $\mathcal{D}(f)$ and $\mathcal{R}(f)$ be the domain and range of $f$ respectively. If Bob uses both his unitary operators from the same set $\mathcal{F}_1$ or $\mathcal{F}_2$ (i.e., Bob's unitary operator $f_1$, $f_2 \Longrightarrow \mathcal{D}(f_1)=\mathcal{D}(f_2)=\mathcal{D}$ (say) and $\mathcal{R}(f_1)=\mathcal{R}(f_2)=\mathcal{R}$ (say), where both $\mathcal{D}$ and $\mathcal{R}$ are either $\varPhi$ or $\varPsi$), then $(N_{A_i},N_{B_i}) \in \mathcal{R} \Longrightarrow~(N_{A_i},M_{B_i}) \in \mathcal{D}$. As Charlie knows exactly the set $\varPhi$ or $\varPsi$ in which the state $(M_{A_i},M_{B_i})$ belongs, thus from the knowledge that $(N_{A_i},M_{B_i}) \in \mathcal{D}$, Charlie gets the information that ``both the bits of Alice's two bits message are equal or not".
	
	Now let the two unitary operators of Bob be $f_1$ and $f_2$, where $f_1 \in \mathcal{F}_1$ and $f_2 \in \mathcal{F}_2$. Then $\mathcal{D}(f_1)=\mathcal{D}(f_2)=\mathcal{D}$ (say) implies $\mathcal{R}(f_1)$ and $\mathcal{R}(f_2)$ are disjoint. Since $\varPhi$ and $\varPsi$ make a partition of the set of all the two qubits Bell states, thus $\mathcal{R}(f_1) \cup \mathcal{R}(f_2)$ contains all the Bell states. As Bob randomly chooses between $f_1$ and $f_2$, therefore from the exact state of $(N_{A_i},N_{B_i})$, Charlie does not know the exact set of the state $(N_{A_i},M_{B_i})$. For example, if Charlie knows $(M_{A_i},M_{B_i})\in \varPhi$, then for Alice's message $b_1b_2$, all the four Bell state can occur as the state of $(N_{A_i},N_{B_i})$. So in this case, the protocol is secure against information leakage.
	
	Hence the collection of all possible choices of Bob's random unitary operators pairs, from the set of Pauli matrices, is $\{(f_1,f_2):  f_1 \in \mathcal{F}_1 \text{ and } f_2 \in \mathcal{F}_2\}$, i.e., there are four options for Bob to choose his pair of unitary operators and they are: $I$ and $\sigma_x$; $I$ and $i\sigma_y$; $\sigma_z$ and $\sigma_x$; $\sigma_z$ and $i\sigma_y$. One can easily check that, if Bob uses any one pair from the above set as his random unitary operators, then both the protocols prevent the information leakage problem.

	\begin{table}[]
		\centering
		\renewcommand*{\arraystretch}{1.7}
		\caption{Different cases of modified MDI-QSDC.}
		\setlength{\tabcolsep}{10pt}
		\resizebox{1.05\textwidth}{!}{
			\begin{tabular}{|c|c|c|c|c|}
				\hline
				\textbf{State of $\mathbf{(M_{A_i},M_{B_i})}$}    & \textbf{Message bits} & \textbf{Alice's unitary}     & \textbf{Bob's unitary} & \textbf{State of $\mathbf{(M_{A_i},M_{B_i})}$} \\ 
				\textbf{ before encoding}    & \textbf{of Alice} & \textbf{operation on $\mathbf{M_{A_i}}$}     & \textbf{operation on $\mathbf{M_{B_i}}$} & \textbf{after encoding} \\ 
				\hline
				\multirow{8}{*}{$\ket{\Phi^{+}}$} & \multirow{2}{*}{00}   & \multirow{2}{*}{$I$}         & $I$                    & $\ket{\Phi^{+}}$                 \\ \cline{4-5} 
				&                       &                              & $\sigma_x$             & $\ket{\Psi^{+}}$                 \\ \cline{2-5} 
				& \multirow{2}{*}{01}   & \multirow{2}{*}{$\sigma_x$}  & $I$                    & $\ket{\Psi^{+}}$                 \\ \cline{4-5} 
				&                       &                              & $\sigma_x$             & $\ket{\Phi^{+}}$                 \\ \cline{2-5} 
				& \multirow{2}{*}{10}   & \multirow{2}{*}{$i\sigma_y$} & $I$                    & $\ket{\Psi^{-}}$                 \\ \cline{4-5} 
				&                       &                              & $\sigma_x$             & $\ket{\Phi^{-}}$                 \\ \cline{2-5} 
				& \multirow{2}{*}{11}   & \multirow{2}{*}{$\sigma_z$}  & $I$                    & $\ket{\Phi^{-}}$                 \\ \cline{4-5} 
				&                       &                              & $\sigma_x$             & $\ket{\Psi^{-}}$                 \\ \hline
				\multirow{8}{*}{$\ket{\Phi^{-}}$} & \multirow{2}{*}{00}   & \multirow{2}{*}{$I$}         & $I$                    & $\ket{\Phi^{-}}$                 \\ \cline{4-5} 
				&                       &                              & $\sigma_x$             & $\ket{\Psi^{-}}$                 \\ \cline{2-5} 
				& \multirow{2}{*}{01}   & \multirow{2}{*}{$\sigma_x$}  & $I$                    & $\ket{\Psi^{-}}$                 \\ \cline{4-5} 
				&                       &                              & $\sigma_x$             & $\ket{\Phi^{-}}$                 \\ \cline{2-5} 
				& \multirow{2}{*}{10}   & \multirow{2}{*}{$i\sigma_y$} & $I$                    & $\ket{\Psi^{+}}$                 \\ \cline{4-5} 
				&                       &                              & $\sigma_x$             & $\ket{\Phi^{+}}$                 \\ \cline{2-5} 
				& \multirow{2}{*}{11}   & \multirow{2}{*}{$\sigma_z$}  & $I$                    & $\ket{\Phi^{+}}$                 \\ \cline{4-5} 
				&                       &                              & $\sigma_x$             & $\ket{\Psi^{+}}$                 \\ \hline
				\multirow{8}{*}{$\ket{\Psi^{+}}$} & \multirow{2}{*}{00}   & \multirow{2}{*}{$I$}         & $I$                    & $\ket{\Psi^{+}}$                 \\ \cline{4-5} 
				&                       &                              & $\sigma_x$             & $\ket{\Phi^{+}}$                 \\ \cline{2-5} 
				& \multirow{2}{*}{01}   & \multirow{2}{*}{$\sigma_x$}  & $I$                    & $\ket{\Phi^{+}}$                 \\ \cline{4-5} 
				&                       &                              & $\sigma_x$             & $\ket{\Psi^{+}}$                 \\ \cline{2-5} 
				& \multirow{2}{*}{10}   & \multirow{2}{*}{$i\sigma_y$} & $I$                    & $\ket{\Phi^{-}}$                 \\ \cline{4-5} 
				&                       &                              & $\sigma_x$             & $\ket{\Psi^{-}}$                 \\ \cline{2-5} 
				& \multirow{2}{*}{11}   & \multirow{2}{*}{$\sigma_z$}  & $I$                    & $\ket{\Psi^{-}}$                 \\ \cline{4-5} 
				&                       &                              & $\sigma_x$             & $\ket{\Phi^{-}}$                 \\ \hline
				\multirow{8}{*}{$\ket{\Psi^{-}}$} & \multirow{2}{*}{00}   & \multirow{2}{*}{$I$}         & $I$                    & $\ket{\Psi^{-}}$                 \\ \cline{4-5} 
				&                       &                              & $\sigma_x$             & $\ket{\Phi^{-}}$                 \\ \cline{2-5} 
				& \multirow{2}{*}{01}   & \multirow{2}{*}{$\sigma_x$}  & $I$                    & $\ket{\Phi^{-}}$                 \\ \cline{4-5} 
				&                       &                              & $\sigma_x$             & $\ket{\Psi^{-}}$                 \\ \cline{2-5} 
				& \multirow{2}{*}{10}   & \multirow{2}{*}{$i\sigma_y$} & $I$                    & $\ket{\Phi^{+}}$                 \\ \cline{4-5} 
				&                       &                              & $\sigma_x$             & $\ket{\Psi^{+}}$                 \\ \cline{2-5} 
				& \multirow{2}{*}{11}   & \multirow{2}{*}{$\sigma_z$}  & $I$                    & $\ket{\Psi^{+}}$                 \\ \cline{4-5} 
				&                       &                              & $\sigma_x$             & $\ket{\Phi^{+}}$                 \\ \hline
			\end{tabular}
		}
		\label{mod_table}
	\end{table}
	
	\section{Conclusion}\label{conclusion}
	In this paper, we have analyzed Niu et al.'s  MDI quantum communication protocols and observed some security issues in both the protocols. We have shown that these protocols are not secure against information leakage, and one bit among two bits of information is always leaked without any active attack. Then we have proposed a modification of these protocols, which are secure against such information leakage problem. We also characterize the set of Pauli operators, which can alternatively be used to bypass the security flaws.
	
	\section*{Authors' note}
	After submitting our current work to arXiv.org (arXiv:2006.05263v1), the authors of Ref.~\cite{niu2018measurement} corrected their flaw independently in Ref.~\cite{niu2020security} by replacing the cover operation from $\{I, \sigma_z\}$ to $\{I, \sigma_x, \sigma_y, \sigma_z\}$. They also simplified the protocol by preparing the EPR pairs all in state $\ket{\psi^-}$. However, in addition to the correction, we also discussed and analyzed the information leakage problem.
	
\bibliographystyle{unsrt}

\end{document}